\documentclass[article,fleqn]{aa}
\usepackage{epsf}
\usepackage{rotate}
\usepackage{times}
\begin{document}
\mathindent1.0cm
\topmargin0.0cm
\thesaurus{06     
 (08.02.3;     
  08.05.3;     
  08.09.3;     
  08.23.1)     
}
\title{The evolution of helium white dwarfs} 
\subtitle{II. Thermal instabilities}
\author{T.\ Driebe\inst{1}, T.\ Bl\"ocker\inst{1},
D.\ Sch\"onberner\inst{2}, and F.\ Herwig\inst{3}}
\offprints{T. Driebe}
\institute{
  Max-Planck-Institut f\"ur Radioastronomie, Auf dem H\"ugel 69, D-53121 Bonn,
  Germany \\ (driebe@speckle.mpifr-bonn.mpg.de; bloecker@speckle.mpifr-bonn.mpg.de)
\and
  Astrophysikalisches Institut Potsdam, An der Sternwarte 16, D-14482 Potsdam,
  Germany \\ 
  (deschoenberner@aip.de)
\and
  Universit\"at Potsdam, 
  Institut f\"ur Physik, Astrophysik, Am Neuen Palais 10,
  D-14469 Potsdam, Germany \\ (fherwig@astro.physik.uni-potsdam.de)
}
\date{Received date /  accepted date}
\maketitle
\markboth{T. Driebe et al.:  Evolution of He-white dwarfs}{}
\begin{abstract}
%
We calculated a grid of evolutionary models for white dwarfs with helium cores
(He-WDs) and investigated the occurrence of hydrogen-shell flashes due to unstable
hydrogen burning via CNO cycling. 
Our calculations show that such thermal instabilities 
are restricted to a certain mass range
($M\approx 0.21\dots 0.30 \,{\rm M}_\odot$),
consistent with earlier studies.
Models within this mass range undergo the more hydrogen shell flashes the less
massive they are. This is caused by the strong dependence of the envelope
mass on the white dwarf core mass. The maximum luminosities from 
hydrogen burning during the flashes are of the order of $10^5\,{\rm L}_\odot$.
Because of the development of a pulse-driven
convection zone whose upper boundary temporarily reaches
the surface layers, the envelope's hydrogen content decreases by 
$\Delta X\approx 0.06$ per flash.

Our study further shows that an additional high mass-loss episode 
during a flash-driven
Roche lobe overflow to the white dwarf's companion does not affect the
final cooling behaviour of the models. Independent of hydrogen shell
flashes the evolution along the final white dwarf cooling branch
is determined by hydrogen burning via pp-reactions down to effective
temperatures as low as $\approx 8000$\,{\rm K}.  

\end{abstract}

%
%
%
\section{Introduction} \label{intro}
%
%
In Driebe et al. (1998) (hereafter referred to as Paper I) we
presented a grid of evolutionary tracks for low-mass white dwarfs
with helium cores (He-WDs) in the mass range 
from $ 0.179$ to $0.414~{\rm M}_{\odot}$.
The lower masses
allow applications to companions of millisecond pulsars. As an example
we derived a cooling age for the He-WD companion of the millisecond pulsar
PSR J1012+5307 of $\sim 6$ Gyr, which is in good agreement with the pulsar's
spin-down age of $\sim 7$ Gyr.
The evolutionary tracks are
based on a $1~{\rm M}_{\odot}$ model sequence extending from the pre-main 
sequence stage through the red-giant branch (RGB) domain.
We forced the models to move off the giant branch and to evolve into the
white dwarf regime by applying large mass-loss rates at appropriate positions
to take into account the binary nature
of He-WDs (for details see Paper I). 

As pointed out in Paper I one of the major results of our study was the
dominant contribution of hydrogen burning to 
the luminosity budget of the He-WDs. Therefore
the final cooling evolution is slowed down, and the derived cooling ages
are notably larger than those found in models which do not consider
nuclear burning or do not find hydrogen burning to be important 
due to much lower envelope masses. 

In the present paper we will discuss the evolution of sequences which
undergo hydrogen shell flashes in detail.
It is organized as follows: In Sect. \ref{instburn} we will briefly
summarize the main reasons for unstable nuclear burning, 
and report in Sect. \ref{insthe} on former studies concerning instabilities in 
He-WDs. In Sect. \ref{calc} we 
describe the evolutionary code used for our calculations.
The main results of the calculations are discussed in Sects. \ref{over} to 
\ref{roche}. Finally, conclusions are given in Sect. \ref{conc}.
\section{Unstable nuclear burning} \label{unstburn}
\subsection{The two general cases} \label{instburn}
There are two reasons for nuclear burning to become unstable: 
The first one 
is the decoupling of the thermal and mechanical structure of
a star due to large electron degeneracy.
If the equation of state gets more and more
independent of the temperature a local increase in energy production
cannot be stabilized by a local expansion with following 
cooling of the affected layers. Therefore a further increase in energy
production will lead to a corresponding increase in temperature
which in turn will again raise the energy release. This phase of
thermally unstable burning will continue until the ongoing temperature
increase results in an effective lifting of degeneracy in the
burning zone. The stronger coupling of thermal and mechanical structure
will then allow for an expansion and the transition to a phase of stable
nuclear burning. This kind of instability is found, for instance, during 
the onset of central helium burning in low-mass stars
($M_{\rm initial}\la 2 \,{\rm M}_\odot, M_{\rm core}
\approx 0.47\,{\rm M}_\odot$ for $Z=0.02$)
on the tip of the red giant branch (central helium flash). 

 
The second reason for thermally unstable nuclear burning is mostly
based on the geometry of the burning region and occurs only during
shell burning. If the shell's mass (its ''thickness``) 
becomes too small compared to its radial extent, the expansion
following a local increase in temperature and energy production 
is insufficient to cool the shell: The thin burning zone
will instead be heated by expansion, and the energy 
production is further increased. 
The thermal runaway only stops when the thickness of the shell
is large enough as to allow for a cooling expansion.
This kind of unstable burning repeatedly takes place
during the double shell-burning phase of AGB stars 
(Schwarzschild \& H\"arm 1965, Weigert 1966).
There, these instabilities are known as
thermal pulses or helium shell flashes caused by thermally
unstable helium burning.

Schwarzschild \& H\"arm (1965)
derived an instability criterion for non-degenerate matter 
by linear perturbation analysis. 
This criterion includes the shell thickness  as well as the 
temperature exponent $\nu$ of the energy generation rate 
( $\epsilon\propto T^\nu$)
as a measure of the temperature dependence of nuclear burning. 
The assumption of non-degeneracy is justified
for the helium layers of AGB models 
during most of the thermal pulse evolution.
According to the Schwarzschild \& H\"arm criterion unstable 
burning is favored when the shell
becomes thinner and the burning is quite temperature sensitive,
as it is the case for helium burning. A similar criterion for the
study of thermal pulses was derived by Sackmann (1977).
In a recent study Frost et al. (1998) report that in advanced
stages of thermal pulse evolution degeneracy in the region
of the helium burning shell may become noticeable and 
can therefore lead to significantly stronger pulses
(degenerate pulses). 
Kippenhahn \& Weigert (1990) give an instability criterion 
which also takes degeneracy into account. 
\subsection{Unstable burning in helium white dwarfs} \label{insthe}
Unlike as in AGB stars, where the helium burning shell becomes thermally
unstable, He-WDs show instabilities related to CNO cycling
which dominates hydrogen burning in 
the lower, i.e. hotter part of the geometrically thin shell. 

Kippenhahn et al. (1967) calculated a 
He-WD model with $M=0.264\, {\rm M}_\odot$ which evolves through
a phase of unstable burning. As a result the track
in the Hertzsprung-Russell diagram (HRD) is reversed and
the He-WD returns to the RGB domain. The evolution of the
white dwarf during the thermal instability was followed in more detail
by Kippenhahn et al. (1968). 
Giannone et al. (1970) found unstable burning in a $0.268\,{\rm M}_\odot$
He-WD model, whereas their models with $M=0.366\,{\rm M}_\odot$ and
$0.426\,{\rm M}_\odot$ did not show any sign of thermal instabilities.
Unlike the previously mentioned studies,
Webbink (1975) found only small hook-like excursions in his He-WD tracks 
close to the point of maximum effective temperature.

Iben \& Tutukov (1986) calculated a $0.298\,{\rm M}_\odot$ He-WD model
based on a $1\,{\rm M}_\odot$ model which suffered from high mass loss
episodes on the RGB, mimicking Roche-lobe overflow to a companion.
They found two strong thermal instabilities to occur
on the cooling branch resulting in a track 
similar to the one of Kippenhahn et al. (1968).

Castellani et al. (1994) calculated a grid of He-WD model sequences for different
metallicities $Z$ using the same technique 
as Iben \& Tutukov (1986). Like Giannone et al. (1970)
they found thermal instabilities only below a certain mass limit
wich depends on $Z$: For $Z=0.01$ and 0.001 no flashes were found for
$M\ga 0.33\, {\rm M}_\odot$ and $M\ga 0.35\, {\rm M}_\odot$, respectively.
For $Z=2\cdot 10^{-4}$ one hydrogen shell flash was found for $M=0.370\, {\rm M}_\odot$
and $M=0.389\, {\rm M}_\odot$.
In a recent study Sarna et al. (1998) found instabilities in their cooling
tracks for $M< 0.2\,{\rm M}_\odot$, quite similar to those of Webbink (1975) who,
however, used rather large time steps. 

Due to large time steps and/or significantly smaller envelope masses 
several studies did not find thermal instabilities at all during the cooling of
He-WDs, as e.\ g.\  Chin \& Stothers (1971), Alberts et al. (1996),
Althaus \& Benvenuto (1997), Benvenuto \& Althaus (1998) and
Hansen \& Phinney (1998).
\section{The evolutionary calculations}  \label{calc}
Besides some minor modifications
we used the evolutionary code described by Bl\"ocker (\cite {Ba}).
Nuclear burning is accounted for
via a nucleosynthesis network including 30 isotopes with all important
reactions up to carbon burning similar as in
El Eid (\cite {El}). The most recent radiative opacities
by Iglesias et al. (\cite{IRW}) and Iglesias \& Rogers (\cite{IR}),
supplemented by those of Alexander \& Ferguson (\cite{AF}) in
the low-temperature region, are employed. Diffusion is not considered.
The initial composition is
$(Y,Z)=(0.28,0.02)$, the mixing length parameter
$\alpha=1.7$ followed from calibrating a solar model.
The Coulomb corrections to the equation of state are those given by
Slattery et al. (\cite{SDW}). 

He-WDs are known to be components in binary systems where
early case B mass transfer (Kippenhahn \& Weigert 1967)
must have taken place during RGB evolution which 
forces the stars to leave the RGB before the onset of helium burning
and to evolve into the white dwarf regime. 
We did not calculate the mass exchange phases during the RGB evolution
in detail
because we are primarily interested in the cooling properties of the
white dwarf
models themselves rather than in the generation of these models by binary star
evolution. Hence we used an approximate approach to get the pre-white dwarf
models (see also Iben \& Tutukov 1986, Castellani et al. 1994, and Paper I):

%
%
%
%
%
%
\begin{figure}[th]
\epsfxsize=8.6cm
\epsfysize=10cm
\epsfbox{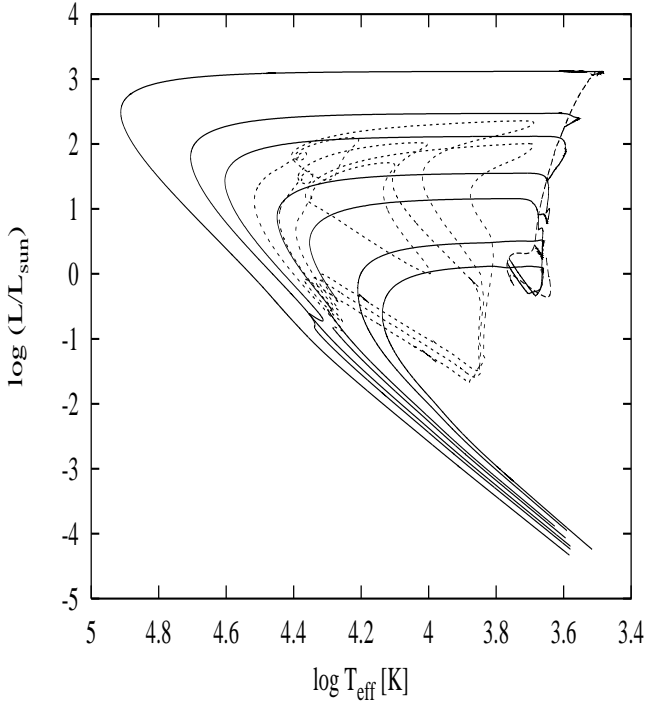}
\caption{\label{oldpic1} Hertzsprung-Russell diagram with evolutionary
tracks for He-WD models. The tracks with solid lines belong to final
white dwarf masses of 0.179, 0.195, 0.234, 0.259, 0.300, 0.331 and
0.414 ${\rm M}_\odot$ (from top to bottom).
The dotted lines show the flash phases for $M=0.234$ 
and $0.259\,{\rm M}_\odot$, the dashed line gives the $1{\rm M}_\odot$
evolutionary path the He-WD models are based upon.
}
\end{figure}

We calculated a 1~${\rm M}_{\odot}$ sequence from the pre-main sequence
phase up to the tip of the RGB.
Along the RGB we applied mass-loss rates $\dot{M}_{\rm R}$ according to
Reimers (\cite{R}) with $\eta = 0.5$.
At appropriate positions high mass loss rates, $\dot{M}_{\rm high}$,
were invoked in order to get models of desired final mass, $M$:
0.179, 0.195, 0.234, 0.259, 0.300, 0.331 and $0.414\,{\rm M}_{\odot}$.
Figure \ref{oldpic1} displays all He-WD sequences in the HRD.
High mass loss varied from $\dot{M}_{\rm high}\approx
10^{-9}~{\rm M}_{\odot}\,{\rm yr}^{-1}$ for  $M \approx 0.15~{\rm M}_{\odot}$ 
to about $10^{-6}\,{\rm M}_{\odot}\,{\rm yr}^{-1}$ for
$M \approx 0.4~{\rm M}_{\odot}$. These values were chosen to
allow the models to hold thermal equilibrium during their further evolution
(with $\dot{M}=\dot{M}_{\rm high}$).
For a more detailed description we refer to Paper I.
Here, we only want to stress the importance of
sufficiently small time steps when the evolution through thermal instabilities
should be followed properly. A brief discussion on this topic is given in 
the Appendix.

\section{Results}    \label{res}
%
%
%
%
\subsection{General remarks}  \label{over}
\begin{figure}[th]
\epsfxsize=8.8cm
\epsfbox{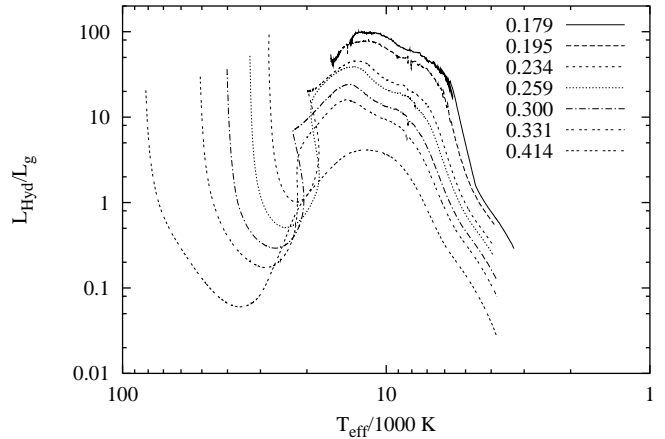}
\caption{\label{oldpic2} 
Ratio of hydrogen ($L_{\rm Hyd}$) to gravothermal luminosity ($L_{\rm g}$)
as a function of $T_{\rm eff}$ for He-WDs of different masses (in ${\rm M}_\odot$).
For the sequences which undergo hydrogen flashes only the final
evolution along the cooling branch is shown.  
}
\end{figure}

\begin{figure}[th]
\epsfxsize=8.8cm
\epsfbox{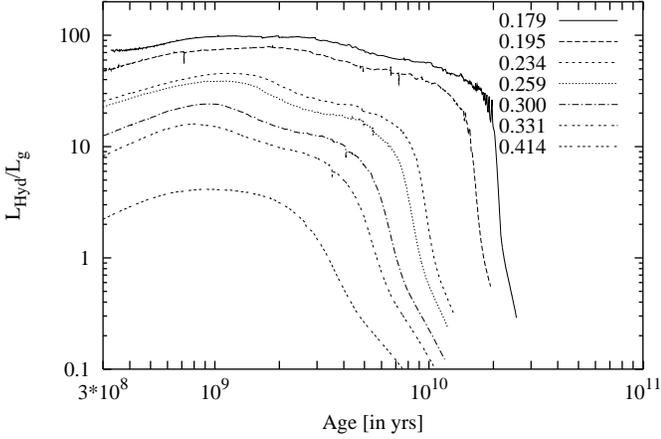}
\caption{\label{oldpic3} 
Same as Fig. \ref{oldpic2}, but as a function of $t_{\rm post-RGB}$. The age at the very 
left of the diagram corresponds to effective temperatures of about $T_{\rm eff}
\approx 15000\dots 18000\,{\rm K}$, the curves end at temperatures of 
$T_{\rm eff}\approx 4000\dots 5000\,{\rm K}$. We adopted $t=0$ when the models
pass  $T_{\rm eff}=10000\,
{\rm K}$ for $M<0.2\,{\rm M}_\odot$ and $T_{\rm eff}=5000\,{\rm K}$ 
for $M>0.2\,{\rm M}_\odot$ after leaving the RGB. 
}
\end{figure}

%
%
We briefly repeat the main results of our calculations which have been
addressed in detail in Paper I. Due to the comparatively large remaining 
envelope masses after termination of the RGB evolution
\begin{table}[h]
\caption[]
{\label{tab1} Total remnant mass $ M$,
mass of the hydrogen-exhausted core $M_{\rm c}$,
total mass of the outer hydrogen layers (''thickness``) $M_{\rm H}$,
envelope mass $M_{\rm env}$, and helium surface
abundance by mass fraction $Y$,
at $T_{\rm eff}=5000$~K for $M>0.2 {\rm M}_{\odot}$ and at 10000 K
for $M\le 0.2 {\rm M}_{\odot}$ after the end of RGB evolution. 
$M_{\rm c}$ is defined by 
the mass coordinate below which $X\le 0.35$ with $X$ being the
mass fraction of hydrogen.
}
\begin{center}
\begin{tabular}{ccrrc}\hline
      \noalign{\smallskip}
$M/{\rm M}_{\odot}$&$M_{\rm c}/{\rm M}_{\odot}$& $\frac{M_{\rm
H}}{10^{-3} {\rm M}_{\odot}}$&$\frac{M_{\rm env}}{10^{-3}
{\rm M}_{\odot}}$& $Y$ 
\\ 
   \noalign{\smallskip}
\hline
0.179&0.1693 &5.061 &10.211 & 0.464\\
0.195&0.1859 &4.937 & 9.598 & 0.461\\
0.234&0.2220 &8.118 &13.098 & 0.354\\
0.259&0.2524 &4.771  &7.232 & 0.312\\
0.300&0.2960 &3.189  &4.746 & 0.301\\  
0.331&0.3281 &2.509  &3.744 & 0.301\\ 
0.414&0.4116 &1.446  &2.175 & 0.301\\
\hline 
\end{tabular}
\end{center}
\end{table}
(see Table \ref{tab1}\footnote{We note that the same table is shown in Paper I
but the data for $M=0.179$ and $0.195\,{\rm M}_\odot$ did not refer
to $T_{\rm eff}=10000\,{\rm K}$ as indicated in the caption but erroneously 
to $T_{\rm eff}=5000\,{\rm K}$ as for the larger masses. }) 
one of the main characteristics of our white dwarf models is that
hydrogen burning due to pp-reactions remains the main energy source
down to effective temperatures well below $10^4\,{\rm K}$ 
(see Paper I and Fig. \ref{oldpic2}).
This residual burning leads to a significant slow-down of the
further evolution, 
resulting in larger cooling ages (typically a few $10^9\,{\rm yr}$, 
see Fig. \ref{oldpic3}) than found for 
He-WD models where hydrogen burning is negligible due to 
smaller, ad-hoc assumed
\footnote{In these calculations the envelope mass is taken as a free parameter
and has not been computed from the mass-loss history of the He-WD progenitor
according to the binary nature of He-WDs.}
envelope masses, or even not considered at all.
The implications of evolutionary envelope masses for the evolution of white 
dwarfs are discussed in Bl\"ocker et al. (1997).

%
%
%
%
%
%
%
%

Besides of this important property of He-WDs and its consequence for 
age determinations of, e.g., millisecond pulsar systems 
as \object{PSR J1012+5307} (cf.\ Paper I), another result of our study
concerns the occurrence of hydrogen shell flashes:
We did not find any unstable hydrogen-burning in the sequences with masses
$M=0.179,\, 0.195$ and $0.414\,{\rm M}_\odot$. 
Only for $M=0.234$ and $0.259\,{\rm M}_\odot$ major hydrogen shell flashes
($L_{\rm Hyd,max}\ga 10^5\,{\rm L}_\odot$)
developed with concomitant extended loops in the HRD.
Only a temporal slight increase of the CNO-luminosity 
on the cooling branch was found in the 
sequences with $M=0.300$ and $0.331\, {\rm M}_\odot$. 

The restricted mass range for the occurrence of hydrogen
flashes agrees with earlier results, e.g. Kippenhahn et al. (1968),
Gianonne et al. (1970) or Castellani et al. (1994). We note that Webbink (1975)
suggested a lower mass limit of $M\approx 0.206\,{\rm M}_\odot$,
in good agreement to our findings. 
The $0.298\,{\rm M}_\odot$ model of Iben \& Tututkov (1986) is fully
unstable, our $0.300\,{\rm M}_\odot$ model only marginally.
The calculations of Castellani et al. (1994) suggest that the upper boundary
for the flash range is higher for lower metallicity. Therefore, their result
of $M^{\rm flash}_{\rm upper}<0.33\,{\rm M}_\odot$ for $Z=0.01$ is consistent 
with our calculations.
Complementary calculations where equilibrium rates for the pp-chains and
the CNO-cycle were used instead of the nuclear network
show three hydrogen shell flashes for $M=0.227\,{\rm M}_\odot$ 
and one strong hydrogen shell flash for $M=0.310\,{\rm M}_\odot$. 
Because we found no hydrogen flashes in our standard
sequences with $M=0.300\,{\rm M}_\odot$ and $M=0.195\,{\rm M}_\odot$
we conclude
that the mass range for the occurrence of instabilities
depends on uncertainties due to different input physics
(e.g. diffusion as in Iben \& Tutukov (1986)), and that the
boundaries $M^{\rm flash}_{\rm lower}=0.21\, {\rm M}_\odot$ and 
$M^{\rm flash}_{\rm upper}=0.30\, {\rm M}_\odot$ may be uncertain by 
about $0.01\,{\rm M}_\odot$.

According to our calculations the number of hydrogen flashes
depends on the mass:
We found one major hydrogen flash
for $M\ga 0.25\,{\rm M}_\odot$, two flashes for $0.25\ga M/{\rm M}_\odot\ga 0.23$
and three flashes for $0.23\ga M/{\rm M}_\odot\ga 0.21$. 
Thus, we conclude that unstable hydrogen burning in He-WDs caused
by the fading CNO cycling (see next section)
is restricted to a certain mass range of
$M\approx 0.21\dots 0.30\,{\rm M}_\odot$. Furthermore, the
number of flashes increases with decreasing white dwarf mass.

It is noteworthy that the cooling properties below $T_{\rm eff}\la 
20000\, {\rm K}$ are independent on the occurrence of hydrogen shell flashes
during the previous evolution. 
Hydrogen burning remains the main energy source
along the cooling branch down to very low effective temperatures
(see Sect. \ref{foof}).
Because He-WDs suffering from hydrogen flashes evolve 
back to the RGB regime one has to account for 
high mass-loss episodes due to Roche lobe overflow
although the time spent away from the cooling branch is small 
compared to the cooling time itself. 
Our calculations show that the cooling is not 
affected by such high mass-loss episodes (see Sect. \ref{roche}).

%
%
%
%
%
\subsection{The main flashes}   \label{tmf}
\begin{figure}[th]
\epsfxsize=8.8cm
\epsfbox{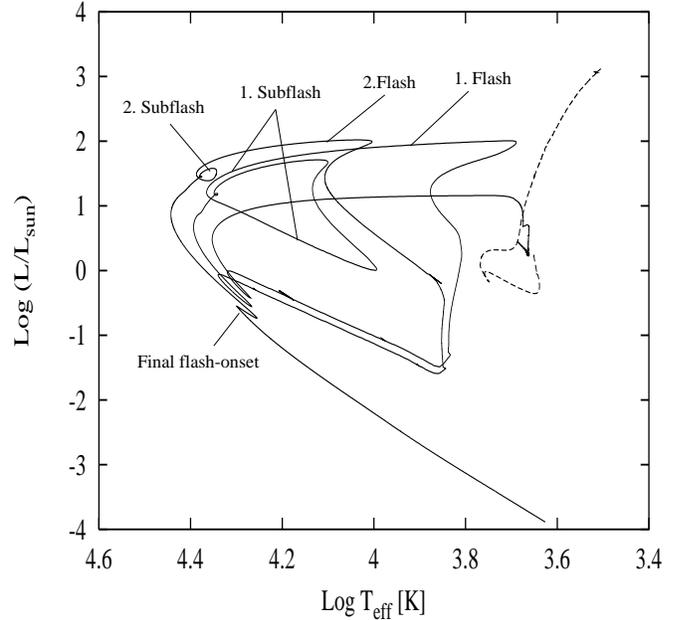}
\caption{\label{pic2}HRD with a complete evolutionary track for the He-WD sequence
with $M=0.234\, {\rm M}_\odot$. The dashed line shows the $1\,{\rm M}_\odot$ track
which was used to generate the post-RGB model. Two main flashes occurred,  
each followed by weaker subflashes 
(see also Fig. \ref{pic4}).}
\end{figure}

\begin{figure}[th]
\epsfxsize=8.8cm
\rotate[r]{
\epsfbox{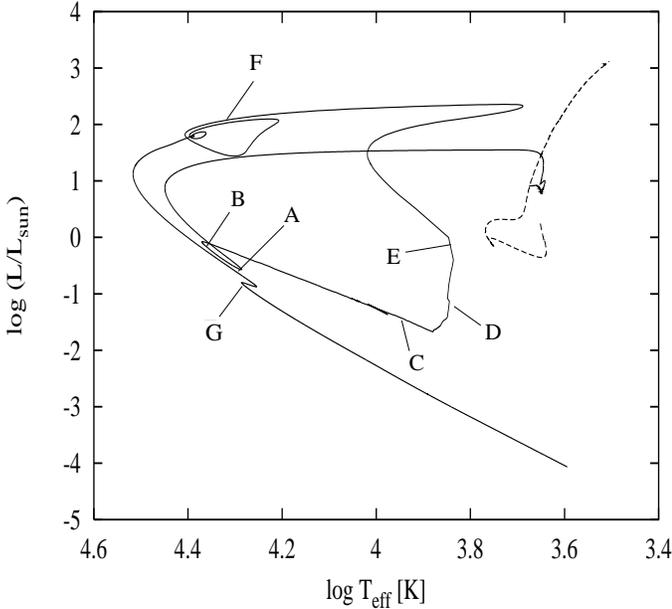}
}
\caption{\label{pic3} Same as Fig. \ref{pic2}, but for the
$M=0.259\, {\rm M}_\odot$ sequence. Here only one major flash was found. 
Properties of the models labeled by capital letters are given in Tab.\ \ref{tab2}.
}
\end{figure}

Figures \ref{pic2} and \ref{pic3} show the complete evolutionary tracks
for the sequences with $M=0.234\,{\rm M}_\odot$ and $0.259\,{\rm M}_\odot$.
%
%
%
%
While the evolution of the  $M=0.259\,{\rm M}_\odot$ model
is characterized by only one major flash resulting in an extended loop in the
HRD (see Fig. \ref{pic3}), the model with $M=0.234\,{\rm M}_\odot$ 
experiences two strong flashes (see Fig. \ref{pic2}).

%
%
%
%
%
%
%
\begin{table}[h]
\caption[]
{\label{tab2}Data for the marked points in Fig. \ref{pic3}. The age was
set to zero  at $T_{\rm eff}=10000\,{\rm K}$ of the post-RGB evolution.
The different evolutionary stages A-G are discussed in the text.
}
\begin{center}
\begin{tabular}{cccc}\hline
      \noalign{\smallskip}
Model&$t_{\rm post-RGB}/10^7\,{\rm yr}$&$\log (T_{\rm eff}/{\rm K})$
&$\log (L/{\rm L}_\odot)$
\\ 
   \noalign{\smallskip}
\hline
\hline
A& 1.469119&4.2865&-0.5726\\
B& 2.803540&4.3678&-0.0754\\
C& 2.803641&3.9532&-1.4070\\
D& 2.803643&3.8488&-1.0786\\
E& 2.803658&3.8413&-0.2007\\
F& 2.803744&4.3272& 2.0633\\
G&13.003598&4.2859&-0.8025\\
\hline
\end{tabular}
\end{center}
\end{table}

After mass loss terminates the RGB evolution, the remnants evolve at almost 
constant surface luminosity to higher effective temperatures towards 
the white dwarf cooling branch.
In this phase the total luminosity is almost
entirely supplied by nuclear burning
due to CNO cycling ($L\approx L_{\rm CNO}$).
The contribution due to contraction is negligible. 
The situation changes when the star reaches the cooling branch. 
Now, the temperature within the hydrogen burning shell becomes 
too low to support further CNO cycling. $L_{\rm CNO}$
drops significantly, and 
contraction sets in until pp-burning takes over the main nuclear energy production. 
Still $L_{\rm Hyd}\gg L_{\rm g}$ holds in this early cooling period
(see e.g.\ Fig. \ref{lt0259} and Fig.~5 in Paper I).
The phase of unstable burning starts 
at typical effective temperatures of $T_{\rm eff}\approx 20000\dots 25000\,{\rm K}$ 
on the cooling branch (close to point A in Fig. \ref{pic3}).
At this stage of evolution the energy production due to CNO cycling rises again 
and finally exceeds the pp-contribution by large amounts.
In Fig. \ref{pic3} this part roughly coincides
with the loop between point A and B in the HRD.
The duration for this period of evolution ($\Delta t_{\rm onset}$)
depends on the remaining envelope mass 
and the thermal timescale of the burning shell,
and ranges from $\Delta t_{\rm onset}\approx 4\cdot10^6\,{\rm yr}$ for 
$M\approx 0.23\, {\rm M}_\odot$
to  $\Delta t_{\rm onset}\approx 4\cdot10^7\,{\rm yr}$ 
for $M\approx 0.30\, {\rm M}_\odot$.
For example, the model with $M=0.259\,{\rm M}_\odot$ gives 
$\Delta t_{\rm onset}\approx 1.7\cdot10^7\,{\rm yr}$, that with 
$M=0.234\,{\rm M}_\odot\, $
only $\Delta t_{\rm onset}\approx 5\cdot10^6\,{\rm yr}$.
At the onset of the second flash for $M=0.234\,{\rm M}_\odot$ one gets
$\Delta t_{\rm onset}\approx 1.2\cdot10^7\,{\rm yr}$
due to the further reduced envelope mass.

The increasing energy release during the flash development causes
a steep temperature gradient in the vicinity of maximum energy production and 
the formation of a pulse-driven convection zone well inside the
hydrogen burning 
shell (beyond point B). The situation is displayed in Fig. \ref{pic5}:
%
%
%
%
%
%
%
%
\begin{figure}[h]
\rotate[r]{
\rotate[r]{
\epsfxsize=8.8cm
\epsfbox{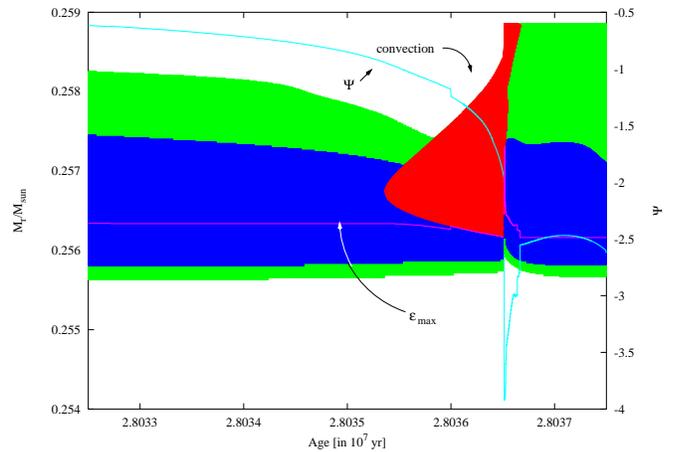}
}
}
\caption{\label{pic5}
Development of the pulse-driven convection zone (intermediate grey shaded region) 
during the hydrogen-flash phase of the $M=0.259\,{\rm M}_\odot$ sequence as a function of time.
Point B from Fig.\ \ref{pic3} is located at $t\approx 2.80354\cdot 10^7$ yr.
Also shown are the lines of maximum energy generation due to hydrogen burning,
$\epsilon_{\rm max}$, the regions where $\epsilon_{\rm max}$ has dropped
to $1 \%$ (dark grey shaded) and $0.1 \%$ (light grey shaded) 
respectively, and the degeneracy parameter $\psi$ 
(right y-axis) at the point with $\epsilon=\epsilon_{\rm max}$.
}
\end{figure}
At $t\approx 2.80357\cdot 10^7\,{\rm yr}$ the convective shell establishes right 
above the locus of maximum energy production due to hydrogen burning. 
The degeneracy inside the burning shell is quite moderate as can be seen 
from the degeneracy parameter $\psi$ at maximum energy generation 
(for the defintion of $\psi$ see e.g. Kippenhahn \& Weigert 1990). 
During the onset of the unstable burning, $\psi\approx -0.5\dots -1$.
 
Once the convection zone is fully established, the evolution is 
rapidly accelerated (roughly between point B and D). 
The typical time scale is now of the order of a few decades. 
The luminosity contribution due to nuclear burning increases several 
orders of magnitude and the convection zone grows
until it extends to the stellar surface
(see Fig. \ref{pic5},
$t\approx 2.80354\dots 2.80364\cdot 10^7\,{\rm yr}$). 
Although $L_{\rm Hyd}$ increases
to about $10^5\,{\rm L}_\odot$ in this phase (near point C in Fig. \ref{pic3}),
the surface luminosity drops by almost
2 orders of magnitude because the increased 
energy production is largely overcompensated
by the energy loss due to the expansion of the envelope. This expansion
roughly doubles the stellar radius and 
leads to a complete lifting of degeneracy in the shell (sharp drop of $\psi$).
For instance, model B has $R\approx 0.05\,{\rm R}_\odot$, and 
beyond point C one finds $R\approx 0.1\,{\rm R}_\odot$.

The maximum hydrogen luminosity reached during the flash is supplied by 
pp-burning, although the onset of the instability is triggered
by CNO cycling (see below, Fig. \ref{pic2a}). For $M=0.259\,{\rm M}_\odot$,
$L_{\rm Hyd, max}\approx 1.3\cdot 10^5\,{\rm L}_\odot$. 
For $M=0.234\,{\rm M}_\odot$ both flashes are of comparable strength with 
$L_{\rm Hyd, max}\approx 2.7\cdot 10^5\,{\rm L}_\odot$ for the first and 
$L_{\rm Hyd, max}\approx 2.3\cdot 10^5\,{\rm L}_\odot$ for the second flash.
The time span between both flashes amounts to $1.8\cdot10^7\,{\rm yr}$
(see also Fig. \ref{pic2a}).

Beyond point C the most luminous part of the flash instability has passed and 
$L_{\rm Hyd}$ decreases while $L_{\rm g}$ increases.
The surface luminosity is again enhanced due to the increase of $L_{\rm g}$.
At point D ($T_{\rm eff}\approx 7000 \,{\rm K}$, see Fig. \ref{pic3}) 
the upper boundary of the pulse-driven convection zone reaches the surface
(at $t\approx 2.80364\cdot 10^7\,{\rm yr}$ in Fig. \ref{pic5}).
Due to the convective transport of hydrogen from the surface layers down to 
the hydrogen burning region the surface hydrogen abundance is
reduced by about $\Delta X\approx 0.06$ when
surface convection establishes.
The small blueward evolution in the evolutionary track
close to point D is related to these composition
changes in the envelope (see Iben \& Tutukov 1986).
The compositional change is consistent with the results of the flash
model of Kippenhahn et al. (1968) who found $\Delta X\approx 0.08$. Iben \& Tutukov
(1986) found a larger amount of $\Delta X\approx 0.2$, probably due to 
their consideration of gravitational and chemical diffusion leading to 
different chemical profiles.
   
Between point D and E in Fig. \ref{pic3} the lower boundary of the pulse-driven
convection moves upwards, and at point E the convection zone (and therefore surface 
convection) vanishes at all. Overall, the convection zone exists for 
$\approx 1200\, {\rm yrs}$, and for $\approx 150$ yrs it extends up to the surface.
Beyond point E the contraction of the inner regions of the hydrogen shell
resumes, while the surface layers react by expansion, resulting in a redward motion 
in the HRD bringing the He-WD almost back to the RGB region. 
Finally, contraction
seizes the surface layers and the star evolves back to higher effective temperatures
towards the cooling branch again.  

Around point F just before re-entering the cooling branch, the so-called 
subflashes develop. Here, contraction initiates another small increase 
of $L_{\rm Hyd}$ to $\approx 30\dots 200\,{\rm L}_\odot$ (compared to 
$\approx 10^5\,{\rm L}_\odot$ during the major flash) leading to a temporary
expansion of the outer layers of the star and only minor 
circle-like excursions in the HRD.   

For illustration, Fig. \ref{pic4} shows the evolution  (arbitrary zero point) 
of the luminosity
contribution due to hydrogen burning, $L_{\rm Hyd}$, and the gravothermal luminosity,
$L_{\rm g}$, during the flash phase of the $M=0.234\, {\rm M}_\odot$ sequence. 
The large plot shows the evolution of the first subflash episode of the first hydrogen
flash phase and the small inlet that of the second phase.  
The peak at $t\approx0.4512\cdot 10^7\,{\rm yr}$ which is not resolved 
within the plot range comes from the major flash, the second peak
at $t\approx0.4518\cdot 10^7\,{\rm yr}$ with 
$L_{\rm Hyd, max}\approx 250\,{\rm L}_\odot$ 
followed by a short period with $L_{\rm g}< 0$, i.e. expansion, marks the subflash. 
This first subflash is then followed by another slight increase in $L_{\rm Hyd}$ at
$t\approx0.4528\cdot 10^7\,{\rm yr}$ with $L_{\rm Hyd, max}\approx 20\,{\rm L}_\odot$ 
causing the little circle in the track at $\log L/{\rm L}_\odot\approx 1.2$
(see Fig. \ref{pic2}).

%
%
%
\begin{figure}[bth]
\epsfxsize=8.8cm
\epsfbox{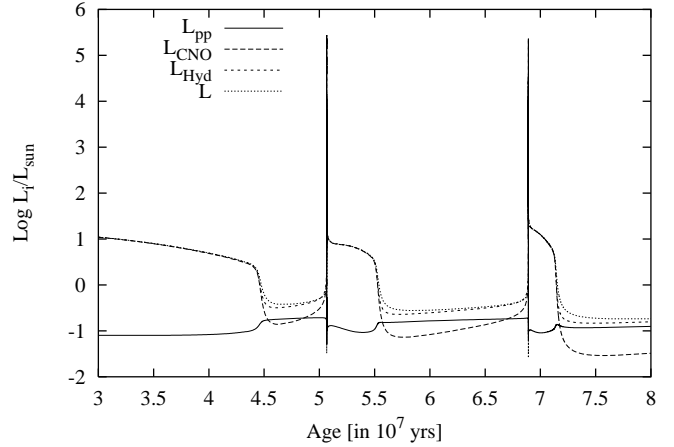}
\caption{\label{pic2a} 
Evolution of surface and hydrogen luminosity, $L$ and $L_{\rm Hyd}$, 
and the contributions due to
CNO cycling and pp-burning as a function of cooling age for
$M=0.234\,{\rm M}_\odot$. The figure shows the age range where 
the hydrogen shell flashes develop. 
Age $t=0$ is given by the first turning point after the track has entered the 
cooling branch (see point A in Fig. \ref{pic3}) and corresponds to a post-RGB age
$t_{\rm post-AGB}\approx 4.62\cdot 10^7$ yr.
}
\end{figure}
%
%
%
%
%
%
%
\begin{figure}[th]
\epsfxsize=8.8cm
\rotate[r]{
\rotate[r]{
\epsfbox{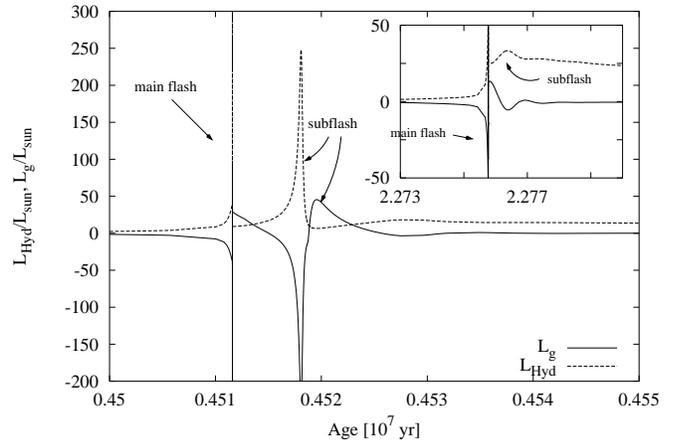}
}
}
\caption{\label{pic4} Evolution of $L_{\rm Hyd}$ (dashed line)
and $L_{\rm g}$ (solid line) as a function
of time for the $0.234\,{\rm M}_\odot$ sequence during flash phase: Large
plot: First flash phase; inlet: second flash phase (same plot labels).
For the definition of $t=0$ see Fig.\ \ref{pic2a}.
}
\end{figure}
%
%
%
After the second major flash (again unresolved in the
time range of this diagram at $t\approx 2.2757\cdot 10^7\,{\rm yr}$) 
a local maximum in $L_{\rm Hyd}$ with 
$L_{\rm Hyd, max}\approx 30\,{\rm L}_\odot$ and slight expansion 
is found at $t\approx 2.2762\cdot 10^7\,{\rm yr}$ (i.e. 
about $5\cdot10^3\,{\rm yr}$ after the main flash)
%
%
%
%
corresponding to the small loop in 
the evolutionary track at $\log L/{\rm L}_\odot\approx 1.5$.

As previously mentioned the instability in He-WDs is caused by
the fading CNO-luminosity on the cooling branch around $T_{\rm eff}\approx 22000\dots
25000\,{\rm K}$. After pp-burning became the dominant contribution of
hydrogen burning, $L_{\rm CNO}$ again increases on a typical 
timescale of a few $10^6$
to $10^7$ yr. Figure \ref{pic2a} shows the situation for the two hydrogen flashes
occurring for $M=0.234\,{\rm M}_\odot$. Beyond $t\approx 4.6\cdot 10^7\,{\rm yr}$
and $t\approx 5.6\cdot 10^7\,{\rm yr}$ $L_{\rm CNO}$, rises again from 
$L_{\rm CNO}\approx 0.1$ to a few $10^2\,{\rm L}_\odot$. With ongoing unstable
burning the outer layers of the He-WD expand and slightly cool the shell
so that $L_{\rm CNO}$ drops again. The peaks in $L_{\rm Hyd}$ ($\approx 10^5\,
{\rm L}_\odot$) are due to pp-burning.

To investigate the dependence of the mass on the occurrence of
hydrogen flashes we applied the criteria 
for unstable burning discussed in Sect. \ref{instburn}.
As already seen in Fig. \ref{pic5} the
degeneracy in the hydrogen shell is quite moderate but far
from being negligible. We used the criterion of 
Kippenhahn \& Weigert (1990) to account for the possible effect
of degeneracy. The criterion states that
nuclear burning in a shell at radius $R_{\rm shell}$ and of thickness
$\Delta R$ becomes unstable if 
\begin{equation} \label{kipcrit}
1-\nabla_{\rm ad}\cdot\frac{4\delta}{4\alpha 
-\frac{R_{\rm shell}}{\Delta R}} > 0\qquad .
\end{equation}
$\nabla_{\rm ad}$ is the adiabatic temperature gradient and $\alpha$
and $\delta$ are given by 
 $\alpha= \left(\frac{\partial \ln \rho}{\partial \ln P}\right)_T$ and 
 $\delta=-\left(\frac{\partial \ln \rho}{\partial \ln T}\right)_P$.
For comparison, we also used the criterion of Schwarzschild \&
H\"arm (1965) which predicts unstable burning if
\begin{equation} \label{schcrit}
\frac{\Delta T}{T_{\rm shell}}>\frac{4}{\nu} \qquad {\rm and} \qquad
\frac{\Delta R}{R_{\rm shell}}<\frac{5}{2}\cdot |Q|
\end{equation}
with $Q\approx-8\dots -4$ and $\Delta T$ being the temperature
difference between lower und upper shell boundary.
A main problem when dealing with these
criteria is the definition of quantities 
as, for example, representative temperatures and pressures in the shell
or the typical radial extent of the shell.

%
%
%
%
\begin{figure}[t]
\epsfxsize=8.8cm
\epsfysize=10.8cm
\epsfbox{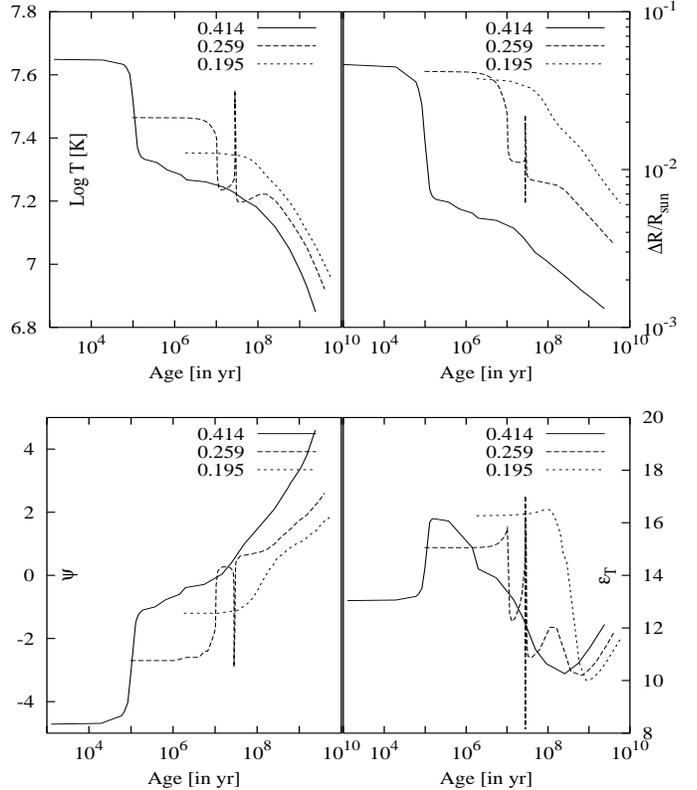}
\caption{\label{pic6} Different quantities as a function of
$t_{\rm post-RGB}$ for the sequences with $M=0.195$ and $0.414\,{\rm M}_\odot$
(no flashes) and $M=0.259\,{\rm M}_\odot$ (one flash):
Upper left:
Temperature at the maximum of energy generation, $\epsilon_{\rm max}$, 
in the hydrogen burning shell. Upper right: Radial extent of the hydrogen
burning shell. Lower left: Degeneracy parameter $\psi$ at 
$\epsilon=\epsilon_{\rm max}$. Lower right: Temperature derivative 
of the energy generation rate, $\epsilon_{\rm T}$, 
at $\epsilon=\epsilon_{\rm max}$.
}
\end{figure}
%
%
%

However, although we find the instability criterions to be fulfilled
in several models the strict application of the aforementioned criteria 
alone cannot explain the restricted mass range for the occurrence of 
hydrogen flashes.
Some general aspects on the model properties important for
the study of unstable burning can be seen from Fig. \ref{pic6} where
different quantities of layers at maximum energy generation
are shown along the cooling branch evolution for the non-flash 
sequences with $M=0.195$ and $0.414\,{\rm M}_\odot$
and the flash sequence with $M=0.259\,{\rm M}_\odot$ covering
most of the He-WD mass range.

As seen before for $M=0.234\,{\rm M}_\odot$, degeneracy of the shell
layers is moderate (lower left panel)
in the upper part of the cooling branch evolution. 
For $M=0.259\,{\rm M}_\odot\,, \psi \approx -0.5\dots 0$ 
at the beginning of unstable burning ($t\approx 2\cdot 10^7\,{\rm yr}$).
For $M=0.414\,{\rm M}_\odot$ degeneracy is slightly lower
($\psi\approx -1$, $t\approx 2\cdot 10^5\,{\rm yr}$)
at the corresponding evolutionary stage,
and for $M=0.195\,{\rm M}_\odot$ slightly higher
($\psi\approx 0\dots +0.5$, $t\approx 5\cdot 10^8\,{\rm yr}$).
For an ideal, non-relativistic gas one obtains for the non-degenerate
limiting case $\alpha=\delta=1$,
whereas strong degeneracy $(\psi\gg 1)\,$
leads to $\alpha\approx \frac{3}{5}(1+\frac{\mu_{\rm e}}{\mu_0}\frac{1}{\psi})\,$ 
and $\,\delta\approx\frac{3}{2}\frac{\mu_{\rm e}}{\mu_0}\frac{1}{\psi}$
with $\mu_0\,$ and $\mu_{\rm e}$
being the mean molecular weights per ion and electron, resp., (see Kippenhahn \& Weigert 1990)
\footnote{ Note that radiation pressure is negligible and $\nabla_{\rm ad}$ does not depend on $\psi$
in the non-relativistic regime.}.
Thus, inspecting Eq.~(\ref{kipcrit}), from $\psi$ alone we would conclude that, if at all, 
degeneracy favours hydrogen shell flashes for lower He-WD masses.

The temperature sensitivity of nuclear burning along the
cooling branch can be seen from the quantity 
$\epsilon_{\rm T}=\partial\ln \epsilon/\partial\ln T (=\nu\, {\rm for}\,\epsilon\propto T^\nu) $
which is shown in the lower right panel of Fig. \ref{pic6}.
When the flash develops for $M=0.259\,{\rm M}_\odot$, one finds
$\epsilon_{\rm T}\approx 12.5$. For the other masses we have
$\epsilon_{\rm T}\approx 15$ ($M=0.414\,{\rm M}_\odot$) and
$\epsilon_{\rm T}\approx 11$ ($M=0.195\,{\rm M}_\odot$).
Because unstable nuclear burning is favoured for higher temperature
sensitivity of the burning (see Eq.\ (\ref{schcrit})) the occurrence 
of flashes is favoured for heavier white dwarfs.
Thus, combining the effects of degeneracy and large
temperature exponents in the energy generation rate might, 
in principle, explain why high and low-mass He-WDs  do not
suffer from hydrogen flashes.

However, it is most likely the radial thickness of the shell
(see upper right panel in Fig. \ref{pic6}) which has the most important
effect on the occurrence of hydrogen flashes.
At the onset of unstable burning we found $\Delta R\approx 10^{-2}\,{\rm R}_\odot$
for $M=0.259\,{\rm M}_\odot$ (the shell's borders are taken at the point with
$\epsilon =10^{-3}\cdot \epsilon_{\rm max}$). 
For $M=0.195\,{\rm M}_\odot$, $\Delta R$ is
approximately a factor of two larger and for $M=0.414\,{\rm M}_\odot$
a factor of two smaller. 
For low-mass He-WDs degeneracy of the shell is of minor importance.
Also, the thickness of the shell is too large as to allow hydrogen shell
flashes.
On the other hand, for high-mass He-WDs the shell is so thin
that the thermal cooling time is smaller than the typical 
time scale for the onset of the instability. Hence, for He-WDs with 
$M\ga0.3\,{\rm M}_\odot$ an instability might be initiated but the 
fast cooling of the shell prevents a hydrogen flash (see also next section). 
%
%
%
\subsection{Final cooling} \label{foof}
\begin{figure}[th]
\epsfxsize=8.8cm
\epsfbox{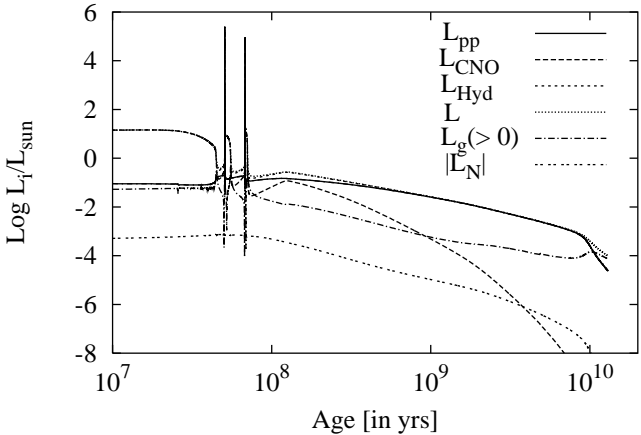}
\caption{\label{lt0234} 
Evolution of different luminosity contributions (see plot labels)
as a function of $t_{\rm post-RGB}$ for the sequence with $M=0.234\,{\rm M}_\odot$:
$L_{\rm pp}$: luminosity due to pp-burning; $L_{\rm CNO}$: luminosity 
due to CNO cycling; $L_{\rm Hyd}$: complete luminosity due to hydrogen 
burning; $L$: surface luminosity; $L_{\rm g}$: contribution of gravothermal
luminosity due to contraction ($L_{\rm g}>0$); $L_{\rm N}$ luminosity contribution 
due to neutrino losses. $t=0$ is adopted as in Fig.\ \ref{oldpic3} 
(see also Fig. \ref{pic2a} for $t=3\dots 8\cdot 10^7 {\rm yr}$).
}
\end{figure}

\begin{figure}[th]
\epsfxsize=8.8cm
\epsfbox{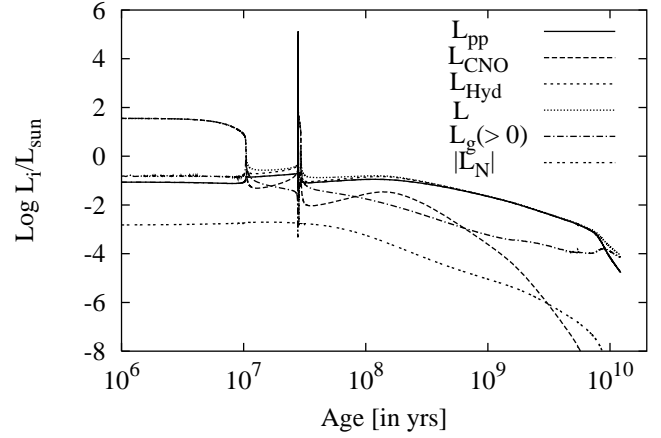}
\caption{\label{lt0259} 
Same as Fig. \ref{lt0234}, but for the $M=0.259\,{\rm M}_\odot$ sequence.
}
\end{figure}

After the He-WD models with $M=0.234\,{\rm M}_\odot$ and $0.259\,{\rm M}_\odot$
have completed their major hydrogen flashes
a final onset of a flash results in a local hook
in the cooling track (see Figs. \ref{pic2} and \ref{pic3}, point G). 
Again this increase
of luminosity due to hydrogen burning is caused by CNO cycling.
This can be seen from Figs. \ref{lt0234} and \ref{lt0259} 
where the evolution of different
luminosity contributions for the two flash sequences are plotted as a function
of $t_{\rm post-RGB}$ for the whole post-RGB evolution.   
%
%
%
%
%
%
%
%
$L_{\rm CNO}$ shows a local maximum at $t_{\rm post-RGB}\approx 1.23\cdot 10^8\,{\rm yr}$
for $M=0.234\,{\rm M}_\odot$,
approximately $5\cdot 10^7\,{\rm yr}$ after the last major flash. 
The corresponding point for $M=0.259\,{\rm M}_\odot$ is at an age of 
$t_{\rm post-RGB}\approx 1.40\cdot 10^8\,{\rm yr}$, which is about  
$10^8\,{\rm yr}$ after the major flash.
At this point of evolution hydrogen shell burning is already dominated by 
pp-burning ($L_{\rm pp}\approx 20\cdot L_{\rm CNO}$),
but the difference between both
contributions is rather small compared to the later evolution
(see Figs.\ \ref{lt0234}
and \ref{lt0259} for $t\ga 1\,$ Gyr). 
Because the characteristic cooling time 
is now smaller than the typical timescale for the onset of 
unstable burning another strong hydrogen flash is prevented. 
Such onsets of unstable
CNO-burning have also been found in the sequences with $0.300\,{\rm M}_\odot$
and $0.331\,{\rm M}_\odot$, resulting in similar hooks on the cooling track (see
Fig.~1 in Paper I).

The further evolution in both flash sequences is comparable to the one
without hydrogen flashes (see for instance Fig.~4 in Paper I). It 
is characterized by quiescent hydrogen shell burning via the pp-chains 
until an age of about 10 Gyr. 
Then hydrogen burning terminates and contraction determines the surface luminosity 
evolution and final cooling (see Figs. \ref{lt0234} and \ref{lt0259}).

Finally, we like to note that
the mass-radius-relation of He-WDs is only affected
by hydrogen shell flashes for $T_{\rm eff}\approx 20000\dots 25000\, {\rm K}$:
The high burning rates associated with the flashes lead to a higher envelope
consumption compared to sequences without flashes,
resulting in slightly smaller
radii and enhanced evolutionary speeds on the upper part of the cooling 
branch. Besides, since the timescale of the 
instabilities is short compared to the characteristic cooling time
(a few $10^7\,{\rm yr}$ compared to Gyr) and most of the time during a flash 
is spent close to the cooling branch,
the influence of hydrogen shell flashes on
the mass-radius-relation is restricted to the onset phase where the changes 
in radius are moderate.
\subsection{Roche lobe overflow during the flashes} \label{roche}
During the hydrogen shell flashes the expansion of the outermost layers
(from a few 0.01 solar radii to a few solar radii) forces the He-WDs 
to evolve back into the RGB domain in the HRD (see Figs. \ref{pic2} 
and \ref{pic3}). Due to the binary nature of He-WDs, in principle,
mass transfer
to a companion (in most cases another white dwarf or a pulsar) should be 
taken into account in this particular evolutionary phase. 
This has been considered, for instance, 
in the calculations of Iben \& Tutukov (1986) leading to a significant removal of 
envelope matter and a faster exhaustion of hydrogen burning.

%
%
%
%
\begin{figure*}[th]
\rotate[r]{
\epsfxsize=11.8cm
\epsfbox{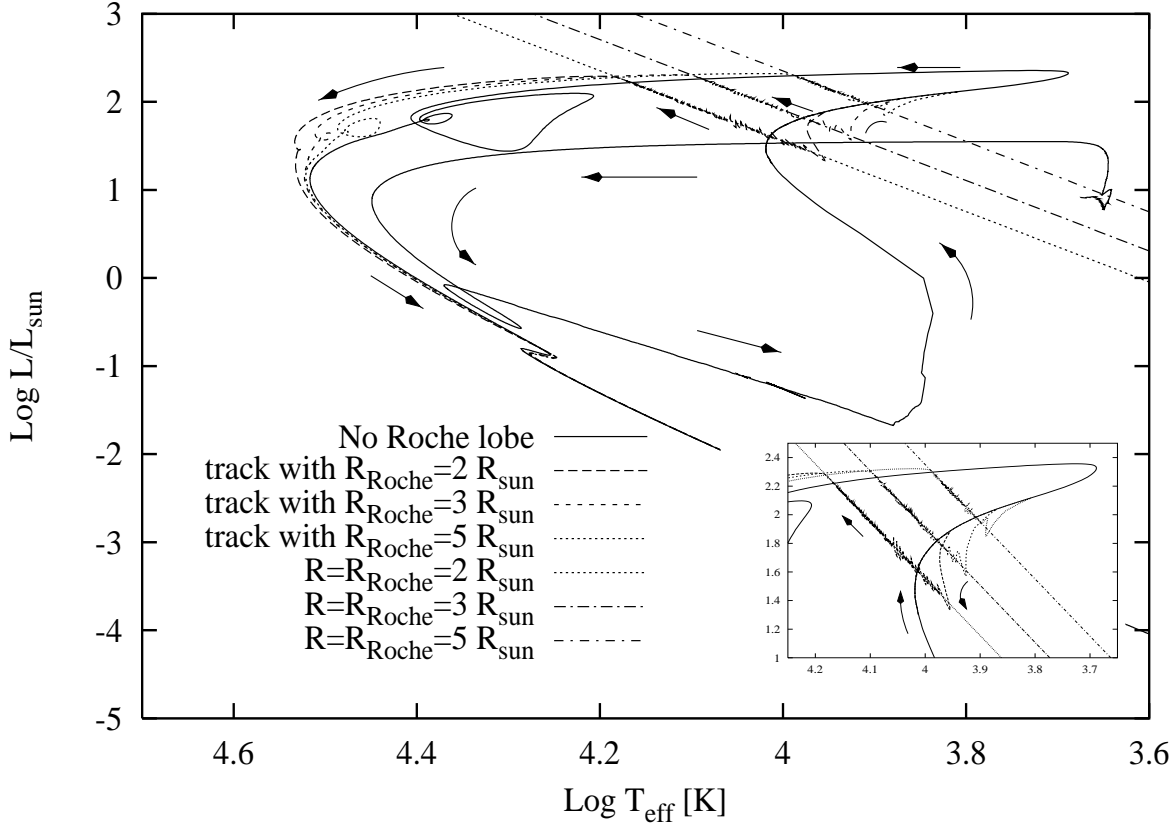}
}
\caption{\label{pic7} HRD with tracks for $M=0.259 {\rm M}_\odot$ and different
Roche radii (2, 3 and 5\, ${\rm R}_\odot$). 
For $R\ge R_{\rm Roche}$ mass loss was adjusted as to provide
$R\approx R_{\rm Roche}$ as long as the star does not evolve back to the white
dwarf domain. Lines of constant radius are also plotted. The arrows indicate
the course of evolution. The inlet in the lower right corner is a blow-up of
the part close to the Roche radii.}
\end{figure*}
Besides the sequences already discussed we thus also 
investigated the evolution of the flash model sequence with $M=0.259\,{\rm M}_\odot$ 
considering high mass-loss episodes due to Roche lobe overflow.
Because we were only interested in the general influence on the final cooling 
properties of our models we used a rather simple algorithm to account for
Roche lobe overflow from the He-WD to a companion: For a given Roche 
radius $R_{\rm Roche}$ (in our case 2, 3 and 5\, ${\rm R}_\odot$) and 
actual model radius $R$, we increased $\eta$ in the Reimers mass loss formula
by a factor of $2\cdot R/R_{\rm Roche}$ for $R>R_{\rm Roche}$. Otherwise, 
$\eta$ was set to its standard value of 0.5 used for all post-RGB calculations.

Figure \ref{pic7} presents the evolutionary tracks in the HRD during the flash for
$M=0.259\,{\rm M}_\odot$ and different assumptions of the Roche radius.
The models quickly reach the line with $R=R_{\rm Roche}$. 
For comparison, also the track without Roche-lobe overflow is shown.
During the adjustment phase,
after the models have passed $R=R_{\rm Roche}$, mass-loss
rapidly rises up to
$\dot{M}_{\rm high}\approx 10^{-3}\,{\rm M}_\odot/{\rm yr}$ 
and forces the models to evolve back to $R\approx R_{\rm Roche}$.
Finally, $\dot{M}\approx 5\cdot 10^{-6}\, {\rm M}_\odot/{\rm yr}$ when
the models evolve along at $R \approx R_{\rm Roche}$.
When Roche lobe overflow ends ($R<R_{\rm Roche}$)  mass loss has
dropped to $\dot{M}=\dot{M}_{\rm Reimers}\approx 10^{-11}\,{\rm M}_\odot/{\rm yr}$ .

After the end of Roche lobe overflow and the subflash phase
the tracks rapidly merge with the one calculated without Roche lobe overflow
at the beginning of final cooling evolution. As Fig. \ref{pic7} shows 
the subflash excursion is the less pronounced the smaller $R_{\rm Roche}$.

\begin{figure}[htb]
\rotate[r]{
\epsfxsize=6.6cm
\epsfbox{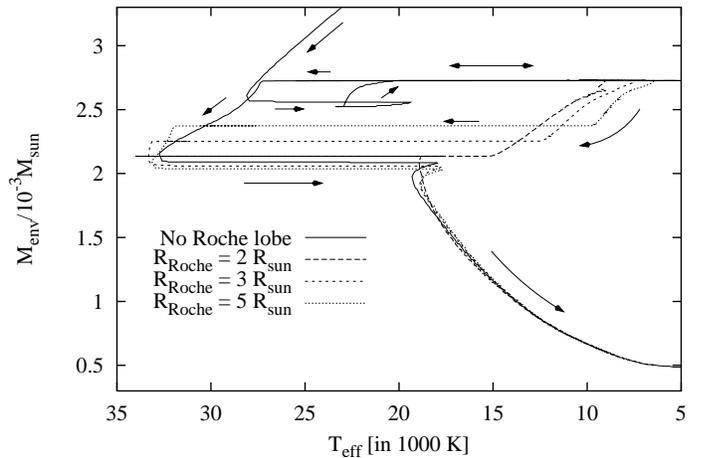}
}
\caption{\label{pic8} Evolution of $M_{\rm env}$ as a function of 
$T_{\rm eff}$ for tracks with $M=0.259 {\rm M}_\odot$ and different 
Roche radii (see labels). The smaller $R_{\rm Roche}$
the faster $M_{\rm env}$ is reduced, but when the star is back on 
the cooling branch the residual envelope mass is almost the same for all Roche 
tracks as it is for the non-Roche lobe track.
}
\end{figure}

\begin{figure}[htb]
\epsfxsize=6.6cm
\rotate[r]{
\epsfbox{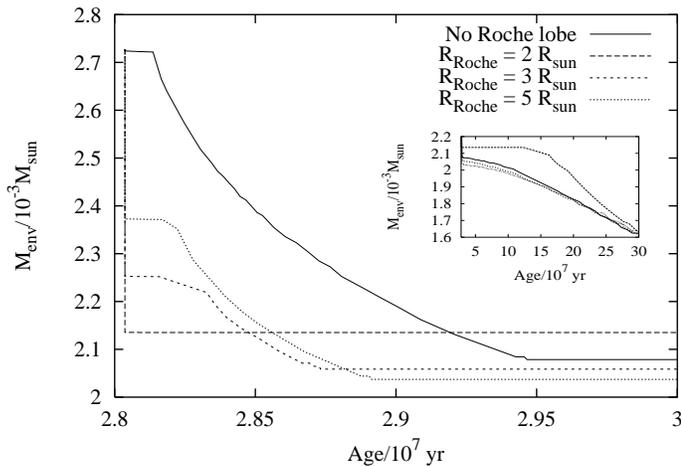}
}
\caption{\label{pic16} Evolution of envelope mass as a function of 
$t_{\rm post-RGB}$ for tracks with $M=0.259 {\rm M}_\odot$ and different 
Roche radii (see labels). The plot shows the reduction of $M_{\rm env}$
during and immediately after Roche-lobe overflow. The inlet shows the evolution
of $M_{\rm env}$ on a larger scale.
}
\end{figure}

\begin{figure}[htb]
\epsfxsize=6.8cm
\rotate[r]{
\epsfbox{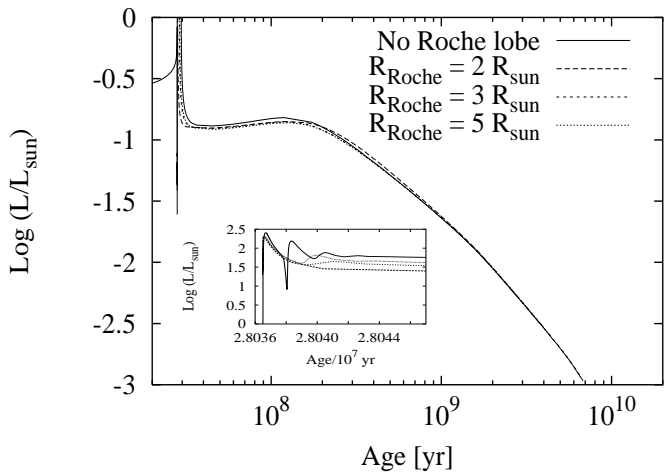}
}
\caption{\label{pic15} Evolution of surface luminosity $L$ as a function of 
$t_{\rm post-RGB}$ for tracks with $M=0.259 {\rm M}_\odot$ and different 
Roche radii (see labels). The models suffering a high mass-loss episode
show the same cooling behaviour as the non-Roche sequence. The inlet shows
the evolution of $L$ during the subflash phase.
}
\end{figure}

The convergence of the tracks shows that the cooling properties of He-WD
models are not seriously affected by the Roche lobe events because 
similar tracks imply similar radii and thus similar envelope masses which
determine the cooling evolution due to hydrogen burning.
This is confirmed by Figs.\ \ref{pic8}, \ref{pic16} and \ref{pic15}.
In Fig.\ \ref{pic8} 
the evolution of the envelope mass, $M_{\rm env}$, is given as a
function of $T_{\rm eff}$ for the tracks from Fig. \ref{pic7}, in
Fig.\ \ref{pic15} the corresponding evolution of surface luminosity as
a function of $t_{\rm post-RGB}$ is plotted. Figure \ref{pic16} shows the reduction
of envelope mass for all sequences as a function of $t_{\rm post-RGB}$
close to $R=R_{\rm Roche}$ and beyond.
Figures \ref{pic8} and \ref{pic16}
indicates that $\sim 6\cdot10^{-4}\,{\rm M}_{\odot}$ of the
envelope mass is lost in the non-Roche model during the flash phase (taken 
from the first local maximum in $T_{\rm eff}$ to the last one). Note that
the temporal increase of $M_{\rm env}$ in Fig.\ \ref{pic8} comes from the
pulse driven convection zone by mixing hydrogen-rich material somewhat below
the layers with $X=0.35$ which determine the border between core and envelope
(cf. Fig.~\ref{pic5}). 

A significant reduction of $M_{\rm env}$
sets in when dominant hydrogen burning has again established at the end
of the subflash phase ( see Fig.~\ref{pic8} at
$T_{\rm eff}\approx 27000\,{\rm K}$
and Fig. \ref{pic16} for $2.81 \la t/10^7\,{\rm yr}\la 2.95$). This envelope
reduction is slowed down when the model enters the cooling branch
(see Fig.~\ref{pic8}
at $T_{\rm eff,max}\approx 33000\,{\rm K}$ and
Fig. \ref{pic16} for $t\ga 2.95\cdot 
10^7\,{\rm yr}$). The envelope is eroded by mass loss and shell burning. 
This can be written as
$\dot{M}_{\rm env}=\dot{M}_{\rm wind}+\dot{M}_{\rm core}$ 
with the mass loss rate $\dot{M}_{\rm wind}$ and the core growth rate
$\dot{M}_{\rm core}$. The core growth rate is given by
$\dot{M}_{\rm core}=\frac{L_{\rm Hyd}}{X\cdot E_{\rm H}}$
with the hydrogen content 
$X$ and the gain of energy due to hydrogen burning per mass unit, 
$E_{\rm H}\approx 6.3\cdot 10^{18}\,{\rm erg}\,{\rm g}^{-1}$.

For sequences without Roche-lobe overflow $\dot{M}_{\rm env}$ is mainly determined 
by shell burning, i.\ e.\ $\dot{M}_{\rm env}\approx \dot{M}_{\rm core}$. 
With a mean value of $L_{\rm Hyd}\approx 30\, {\rm L}_\odot$ and $X\approx 0.7$ it is 
\hbox{$\dot{M}_{\rm core}\approx 4.3\cdot 10^{-10}\,{\rm M}_\odot\,{\rm yr}^{-1}$.} 
With $\Delta t\approx 1.3\cdot 10^6\,{\rm yr}$
(see  Fig. \ref{pic16}) one obtains
$\Delta M_{\rm env}\approx \dot{M}_{\rm core}\cdot 
\Delta t\approx 5.6\cdot 10^{-4}\,{\rm M}_\odot$ 
in good agreement with the value obtained from Figs. ~\ref{pic8} and \ref{pic16}.

For sequences with Roche-lobe overflow the situation is different. 
Here, the consumption of envelope mass occurs preferentielly 
due to the high mass-loss phase when the models are close to their Roche limit
($T_{\rm eff}\approx 8000\dots 15000\,{\rm K}$, see Fig.\ \ref{pic8}).
For instance, for the sequence with $R_{\rm Roche}=3\,{\rm R}_\odot$
it is $\Delta M_{\rm env}\approx 4.5\cdot 10^{-4}\,{\rm M}_\odot$, 
corresponding to a mean mass loss of
$\dot{M}_{\rm wind}\approx 10^{-5}\,{\rm M}_\odot\,{\rm yr}^{-1}$  
lasting for $\Delta t\approx 40\,{\rm yr}$.
Afterwards $M_{\rm env}$ is further reduced
by \hbox{$\Delta M_{\rm env}\approx 2\cdot 10^{-4}\,{\rm M}_\odot$} 
due to nuclear burning (see Fig.\ \ref{pic16}).
%
%
%
%
%
%
%
%
%
%
%
%
%
%
%
%
%
%
%
%
%
%

In the end, the envelope masses are almost the same when the different models
enter the final cooling branch ($M_{\rm env}\approx 2.05\cdot10^{-3}\,{\rm M}_{\odot}$, 
see Fig.\ \ref{pic16}), i.e. the residual envelope mass and the corresponding
cooling evolution are rather independent from the evolutionary history.

On one hand, the latter result is consistent with the results of Iben \& Tutukov (1986). 
In the case of Roche lobe overflow the 
high mass loss determines the evolutionary speed, whereas otherwise
the evolution is characterized by the thermal timescale of the envelope.
On the other hand, in their $0.298\,{\rm M}_\odot$ model the high mass loss 
causes a considerable 
loss of envelope mass ($\approx 1.5\cdot 10^{-3}\,{\rm M}_\odot$) and
prevents hydrogen burning from becoming ever dominant again. Consequently, their model
cools down quite rapidly. The cooling time is of the order of a 
few $10^8\,{\rm yr}$ whereas our models are
controlled by residual hydrogen burning leading to cooling times 
of the order of Gyr (see Figs. \ref{lt0234} and \ref{lt0259}).

Thus, from our calculations it appears that different mass loss histories during the 
phase of the hydrogen shell flashes do not influence the final
cooling of He-WDs significantly. This result seems to be consistent 
with the calculations of Kippenhahn et al. (1968): Their  $M=0.264\,{\rm M_\odot}$
model evolves through one major hydrogen shell flash and one strong subflash,
both bringing the white dwarf radius above the critical
Roche radius (points Q to R
and T to U in their Fig.~1), resulting in a total mass loss of 
about $10^{-3}\,{\rm M}_\odot$ which is comparable to
the $6\cdot 10^{-4}\,{\rm M}_\odot$ of the present calculation 
with only one high mass-loss episode. 
After a rapid evolution back to the cooling branch 
the evolution is slowed down, and
at $\log (L/{\rm L}_\odot)\approx -2$ (point W in their calculation) 
a cooling age of approximately 2.55 Gyr is reached, comparable with 
our $M=0.259\,{\rm M}_\odot$ model which has at the same position
$t_{\rm post-RGB}\approx 1.85\,$ Gyr.
On the other hand, by assuming a linear correlation
between $\log L/{\rm L}_\odot$ and $\log t_{\rm cool}$ as a first
approximation, the cooling age of the model of Iben \& Tutukov (1986)
can be estimated to be only $t_{\rm cool}\approx 2.3\cdot 10^8\,{\rm yr}$
at $\log (L/{\rm L}_\odot)\approx -2$ (using their points S and T, see their
Fig.~1 and Table 1).  This value is almost one order of magnitude below
the one given by models which allow hydrogen burning to continue.


%
%
%
\section{Summary}  \label{conc}
We have calculated evolutionary models of low-mass white dwarfs with
helium cores to study in detail their cooling behaviour and their evolution 
including phases of thermally unstable hydrogen burning (hydrogen shell
flashes). We found that the occurrence of these thermal instabilities is 
restricted to the mass range $0.21\la M/{\rm M}_\odot\la 0.3$.
It is noteworthy that a sufficient temporal resolution is essential to
avoid numerical fluctuations in the energy output of the shell during
early cooling branch evolution.

The flashes occur during the fast cooling of the shell
at the beginning of the cooling branch evolution. This especially affects
the lower part of the shell region where CNO cycling is the
dominant contribution to hydrogen burning. 
CNO-burning is temperature sensitive enough to cause thermal instabilities
if two conditions are fulfilled: The shell is thin enough (this is obviously 
not the case for $M/{\rm M}_\odot \la 0.21$)
and the cooling time is large enough as to avoid an
extinction of the shell before the instability is fully established 
(this does not hold for $M/{\rm M}_\odot\ga 0.3$). The final cooling 
is not affected from possible flash events, i.e. hydrogen shell burning
establishes again as dominant energy source at the end of the flash episode as 
in the case of non-flash models.
   
This result also holds when Roche-lobe overflow of the He-WD to a companion is 
considered during the expansion into the RGB regime.
One of the main features related
to hydrogen shell flashes is the change of envelope composition due to a pulse-driven
convection zone which temporarily extends to the outermost layers of the star. 
The resulting drop in hydrogen is $\Delta X\approx 0.06$.

While the typical duration of a hydrogen flash is short compared to 
the cooling time of He-WDs ($10^7\,{\rm yr}$ compared to Gyr) and He-WDs 
evolve close to their cooling tracks for most of the time during a flash, the 
influence of these unstable phases on the mass-radius-relation of helium white 
dwarfs is moderate and restricted to effective temperatures of $20000\dots 25000$ K.

\begin{acknowledgements}
F.H. acknowledges funding by the {\em Deutsche
Forschungsgemeinschaft} (grant La 587/16).
\end{acknowledgements}
%
%
%
%

%
%
%
%
\begin{appendix}
\section{Remarks on the appropriate temporal resolution}\label{app}
The question of sufficient resolution always arises
when one deals with instabilities in evolutionary models. As mentioned in
Sect. \ref{insthe} thermal instabilities in He-WD model calculations 
can obviously be missed if too large time steps are used. 
Large time steps may also cause fluctuations in the shell energy output. 

%
%
%
%
%
\begin{figure}[ht]
\epsfxsize=8.6cm
\epsfysize=10cm
\epsfbox{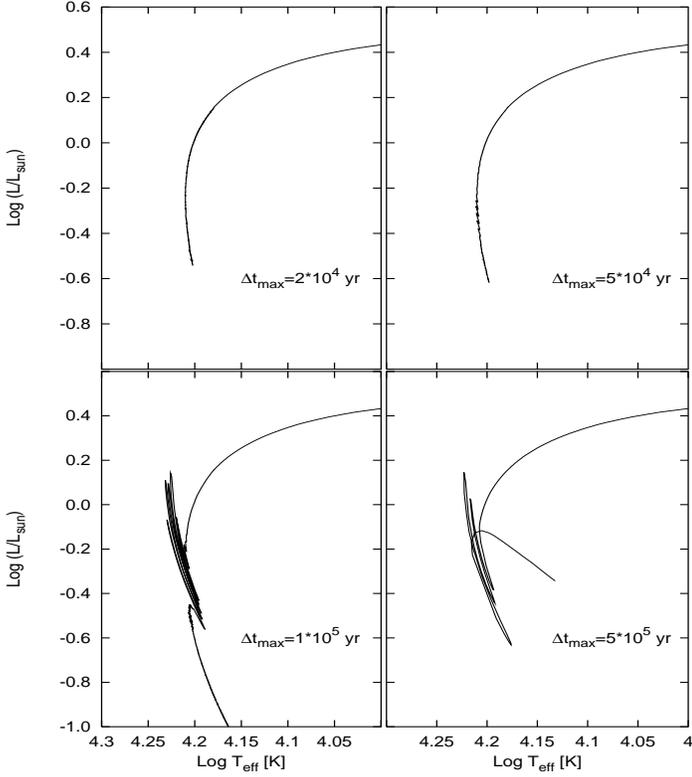}
\caption{\label{pic0a} Hertzsprung-Russell diagram
for an evolutionary track with $M= 0.195\,{\rm M}_\odot$
and different temporal resolutions (see labels). The upper left panel shows
the final track with the appropriate resolution giving a
smooth course of the hydrogen luminosity contribution $L_{\rm Hyd}$ 
(see Fig. \ref{pic0b}). For the other tracks deviations caused by 
local fluctuations in $L_{\rm Hyd}$ (see Fig. \ref{pic0b}) 
increase with increasing maximum allowed time step.}
\end{figure}

\begin{figure}[ht]
\epsfxsize=8.6cm
\epsfysize=10cm
\epsfbox{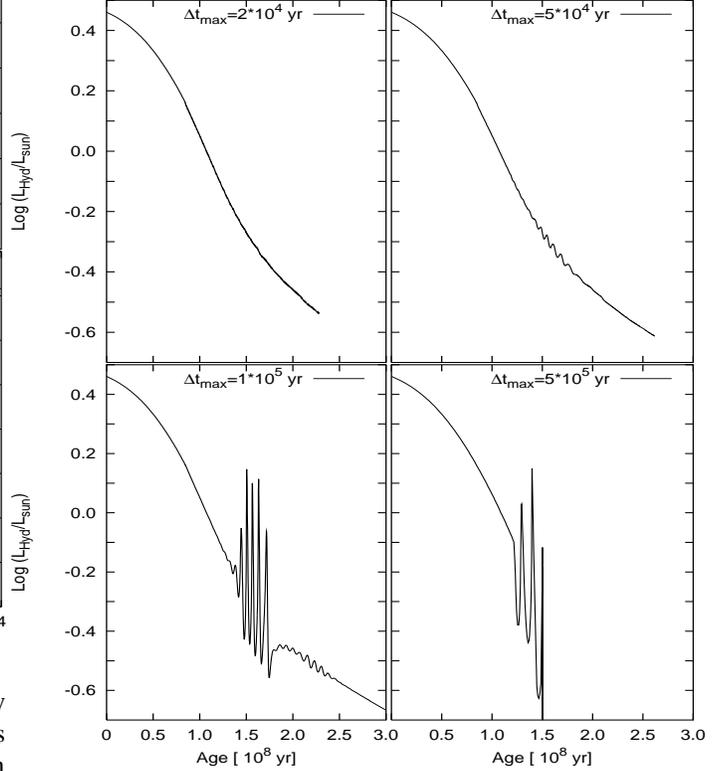}
\caption{\label{pic0b}
Evolution of the luminosity contribution due to hydrogen shell
burning ($L_{\rm Hyd}$) for $M= 0.195\, {\rm M}_\odot$ with
different temporal resolution (see also Fig. \ref{pic0a}).}
\end{figure}

The four panels in Fig. \ref{pic0a} show a part of the evolutionary track
of our He-WD model with $M=0.195\, {\rm M}_\odot$ in the vicinity of the
turn-around point, $T_{\rm eff}=T_{\rm eff, max}$
with different temporal resolution.
The corresponding evolution of the luminosity contribution due to 
hydrogen shell
burning, $L_{\rm Hyd}$, is shown in the four panels of Fig. \ref{pic0b}.

The track with sufficient temporal resolution (upper left panel in Fig. 
\ref{pic0a}, $\Delta t_{\rm max}=2\cdot 10^4 {\rm yr}$)
shows no perturbations due to fluctuations in the shell energy production 
(smooth curve in Figs. \ref{pic0a} and \ref{pic0b}). When the maximum time step
is increased to $\Delta t_{\rm max}=5\cdot 10^4 {\rm yr}$ small perturbations 
occur in $L_{\rm Hyd}$ and in the track (for $-0.4\la \log L/L\odot\la -0.25$,
see upper right panel in Fig. \ref{pic0a}).
When the maximum time step is further increased by a factor of 2 (lower left panel
in Figs. \ref{pic0a} and \ref{pic0b}) the fluctuations in $L_{\rm Hyd}$ and 
in the track are much more prominent but the evolution stabilizes after a 
final hook in the track at $\log L/{\rm L}_\odot\approx -0.5$. 
In this case the shape of the tracks corresponds with 
those of e.g. Webbink (1975) or Sarna et al. (1998).
Finally, for $\Delta t_{\rm max}=5\cdot 10^5 {\rm yr}$ 
the calculations became numerically unstable.

Therefore, we selected $\Delta t_{\rm max}\la 2\cdot 10^4 {\rm yr}$ as the maximum time step
for our calculations.
Additionally, to handle the large local changes in the luminosity budget of the
models we coupled the evolutionary time step to changes in $L_{\rm Hyd}$ by reducing
$\Delta t_{\rm max}$ by a factor of 2 if $L_{\rm Hyd}$ changes by more than 5 \%.
%
%
%
%
%
%
%
%
%
%
\end{appendix}
\end{document}